# Number of wireless sensors needed to detect a wildfire

Pablo I. Fierens

Instituto Tecnológico de Buenos Aires (ITBA)

Physics and Mathematics Department

Av. Madero 399, Buenos Aires, (C1106ACD) Argentina

pfierens@itba.edu.ar

**Abstract**

The lack of extensive research in the application of inexpensive wireless sensor nodes for the early detection of wildfires motivated us to investigate the cost of such a network. As a first step, in this paper we present several results which relate the time to detection and the burned area to the number of sensor nodes in the region which is protected. We prove that the probability distribution of the burned area at the moment of detection is approximately exponential, given that some hypotheses hold: the positions of the sensor nodes are independent random variables uniformly distributed and the number of sensor nodes is large. This conclusion depends neither on the number of ignition points nor on the propagation model of the fire.

# 1 Introduction

The well-established literature on wireless sensor networks continuously mentions prevention and early detection of wildfires as a typical application of the field. However, to the best of our knowledge, only a few proposals of inexpensive sensor networks have been actually made in the literature (e.g., Rodriguez et al. (2000); Yu et al. (2005)) and only one has been implemented and tried to a small scale (Chen et al. (2003); Doolin et al. (2004); Glaser (2004); Doolin and Sitara (2005)).

The lack of actual experiences on the implementation of wireless sensor networks for the detection of wildfires and the current problematic of forest fires in Argentina lead a group of researchers at ITBA to become involved in a mid-term project for the development of such a network. As part of the project, forestry companies were consulted about their needs and experiences on fire detection. Companies in the region use mainly two alternative ways of fire detection (private communication). On one hand, the most extended practice is the visual inspection of large areas (with a coverage radius of up to 20 km) from high towers and the daily walk of personnel through pre-established paths during the fire-season. This type of system is very cheap because its main cost is represented by the low wages of the few people involved in the direct observation. On the other hand, a few companies have also implemented the observation through cameras in the visual and infrared ranges. However, this class of system is usually considered too expensive because the relatively high initial cost of installation of the infrared cameras.

Under this situation, several companies were interested in the idea of a wireless sensor network for the detection of wildfires, but they were also concerned on the cost of the system per unit of area. This problem can be decomposed mainly into two parts: a) the cost of each individual node; b) the number of nodes which are needed per unit of area (i.e., the number of nodes per squared meter). In this paper we investigate part b), while part a) will be presented elsewhere.

## 1.1 Variables under analysis

We shall work with simple two dimensional models and we shall deal only with propagation of surface fires. Moreover, we shall keep the model of the wireless sensors as simple as possible, that is, we shall assume that all sensors allow for the detection of the fire as soon as the fire reaches their location. Finally, we shall not concern ourselves with the difficulties of the communication among the sensors, which may be impaired by the activity of the fire itself (Heron and Mphale (2004), Mphale et al. (2007)).

Under this setting, there are many variables related to the number of wireless sensors that can be studied. We choose to of them, the *time to detection* ($T_d$) and the *area already burnt at the time of the detection* ($A_d$). Both variables can be related, in turn, to the resources needed for contention of the fire after it has been detected.

Since the number of wireless sensors may vary according to the extension of the region that must be protected, we shall use the *characteristic distance between sensors* ($D$) as reference independent of the actual area. Although the definition of the characteristic distance between sensors may vary slightly from one setting to another, in all cases it subsumes under a single value how widely spaced the sensors are.

The rest of the paper is structured as follows. In Section 2, we present some simple results for the case where the nodes are distributed in a regular pattern across the area of interest. In Section 3, we analyze the expected time to detection and the burned area before detection when the sensor nodes are randomly distributed in the protected area. Section 4 summarizes the main conclusions of the paper and mentions some ideas for future work.

## 2 Regularly distributed sensors

In this section, we analyze the case where the sensors are located in a regular grid in such a way that the distance between any pair of them in the same row or the same column is $D$, the characteristic distance in this setting. As a further simplification, we shall assume that the region to be protected is a rectangle whose sides are integer multiples of $D$. It is easy to extend the work in this section to more general regions by partitioning them into small rectangular pieces.

### 2.1 Circular propagation at constant rate

We assume that surface fires propagate at a constant rate of spread ($R$) in all directions and that the probability of ignition is uniformly distributed inside the protected area. Both the uniform distribution of the probability of ignition and the fact that the nodes are located in a regular lattice enable us to reduce the study of the detection of a fire to a much smaller area corresponding to a square delimited by four sensor nodes, one on each vertex. Furthermore, this square can be split up into four smaller squares, as shown in Figure 1. Given the assumption that a fire propagates in all directions at the same speed, if a fire originates in Region $i$ ($i$ =1, 2, 3, 4 – see Figure 1), it will be detected by sensor $i$ first. Therefore, we can further limit our analysis to only one of the four regions and the corresponding node, say, Region 1 and sensor node 1. In other words,

$$P(T_d \leq x) = P[T_d \leq x | \text{the fire originated in Region 1}]$$

Notice that $T_d$ is simply the distance from the ignition point to the sensor node divided by the constant rate of spread $R$. Since the distribution of the ignition point is uniform inside Region 1, we have

$$P[T_d \leq x | \text{the fire originated in Region 1}] =$$

$$= P[\text{Distance from sensor 1 to the ignition point} \leq Rx | \text{the fire originated in Region 1}] =$$

$$= \frac{\text{Area of the surface whose points are at a distance} \leq Rx \text{ of sensor 1}}{\text{Area of Region 1}}.$$

After some work, the latter expression can be found to be

$$P(T_d \leq x) = \frac{\pi}{4} \cdot \left(\frac{Rx}{D/2}\right) \text{ si } Rx \leq D/2, \quad (1)$$

$$P(T_d \leq x) = \left[\frac{\pi}{4} - \tan^{-1}\sqrt{\left(\frac{Rx}{D/2}\right)^2 - 1}\right] \cdot \left(\frac{Rx}{D/2}\right)^2 + \sqrt{\left(\frac{Rx}{D/2}\right)^2 - 1} \text{ si } D/2 \leq Rx \leq D/\sqrt{2}, \quad (2)$$

$$P(T_d \leq x) = 1 \text{ si } Rx \geq D/\sqrt{2}. \quad (3)$$

Further algebraic work leads to

$$P\left(T_d \geq \frac{D}{2R}\right) = 1 - \frac{\pi}{4} \approx 0.215, \quad (4)$$

$$E[T_d] = \frac{\sqrt{2} + \log(1+\sqrt{2})}{6} \frac{D}{R} \approx 0.3826 \cdot \frac{D}{R}, \quad (5)$$

$$E[T_d^2] = \frac{1}{6}\left(\frac{D}{R}\right)^2, \quad \text{var}(T_d) = \frac{6 - (\sqrt{2} + \log(1+\sqrt{2}))^2}{36}\left(\frac{D}{R}\right)^2 \approx 0.0203 \cdot \left(\frac{D}{R}\right)^2, \quad (6)$$

$$E[A_d] = \frac{\pi}{6} D^2 \approx 0.52 \cdot D^2. \quad (7)$$

Some comments are due. Equation (4) points out that more than 80% of fires will be detected in less than $D/2R$ and, hence, their area will be smaller than $\pi D^2/4$. Equation (5) says that, as expected, the mean detection time is proportional to the ratio $D/R$. Finally, while Equations (1)-(3) imply that no fire will have an area greater than $\pi D^2/2$ at the moment of detection, Equation (7) says that the expected value of such area is approximately $D^2/2$. In particular, note that the expected value of the area of the fire at the moment of detection *does not depend on the rate of spread of the fire, but it only depends on the characteristic distance D*.

## 3 Randomly distributed sensors

In this section, we analyze the case where the sensors are randomly distributed in the region to be protected. In this case, the characteristic distance $D$ is the mean distance between sensors computed as

$$D = \sqrt{\frac{A}{N}}, \qquad (14)$$

where $A$ is the total area of the protected region and $N$ is the total number of sensors.

We shall first show that, when the number of sensors is large, the burned area at the moment of detection $A_d$ *has a simple probability distribution which is independent of the propagation model*. We shall then show simple approximations for the statistics of the time to detection for simple propagation models.

### 3.1 Probability distribution of the burned area

The probability that the burned area $A_d$ is greater than a given number $x$ is equal to the probability that the fire "has not found" any sensor inside the burned region. Since sensor nodes are randomly distributed across the protected region, and assuming that the position of each sensor is independent of that of the others, we have

$$P(A_d > x) = \left(1 - \frac{x}{A}\right)^N,$$

where $A$ is the total area of the protected region and $N$ is the total number of sensor nodes. Using Equation (14), we can write

$$P(A_d > x) = \left(1 - \frac{x}{ND^2}\right)^N = \left(1 - \frac{x/D^2}{N}\right)^N.$$

Note that, as $N \to \infty$,

$$P(A_d > x) = \left(1 - \frac{x/D^2}{N}\right)^N \xrightarrow[N \to \infty]{} \exp\left(-\frac{x}{D^2}\right).$$

Formally, we have the following

**Theorem 1.** *Assume the locations of sensor nodes are independent and identically distributed random variables with uniform distribution across the protected region. Furthermore, assume that the position of the sensor nodes and the ignition points (there may be more than one) are independent random variables. Let N be the number of sensor nodes and A(N) the surface area of the protected region, which varies with N in such a way that A(N) = D²N, where D is a positive constant. Then, as the number N of sensor nodes goes to infinity, the random variable $A_d$ corresponding to the burned area at the moment of detection converges in distribution to an exponential random variable with parameter λ=1/D².*

Probably, the most salient feature of Theorem 1 is that *it does not depend on the propagation law of the fire and it depends neither on the number nor on the distribution of the ignition points.* In other words, Theorem 1 states that, if the protected area is large and the number of sensor nodes is also large, then

$$P(A_d > x) \approx \exp\left(-\frac{x}{D^2}\right), \quad \mathrm{E}[A_d] \approx D^2, \quad \mathrm{var}(A_d) \approx D^2.$$

## 3.2 Probability distribution of the time to detection

Theorem 1 leads to the following simple

**Corollary 1.** *Assume that there is a deterministic law F such that, at each time instant t, F(t) represents the total burned area at time t. If the hypotheses of Theorem 1 are satisfied, then, as the number of sensor nodes N goes to infinity,*

$$P(T_d > t) \xrightarrow[N \to \infty]{} \exp\left(-\frac{F(t)}{D^2}\right).$$

In the following paragraphs, we consider two simple examples of application of Corollary 1.

### 3.2.1 Circular propagation at constant rate

Note that, as the surface area of the protected region increases, we may ignore the cases where the fire develops near the borders of the region. Then, it is easy to see that the function *F* in Corollary 1 for the case of circular propagation at constant rate of spread *R* is given by

$$F(t) = \pi \cdot (Rt)^2.$$

Then, for a large area covered with a large number of sensors, we can make the following approximation:

$$P(T_d > t) \approx \exp\left(-\frac{\pi R^2 t^2}{D^2}\right).$$

So we can estimate the expected value of the time to detection as

$$E[T_d] = \int_0^\infty P(T_d > t)\,dt \approx \int_0^\infty \exp\left(-\frac{\pi R^2 t^2}{D^2}\right)dt = \frac{D}{2R},$$

where we have omitted the details of the calculation. In a similar fashion, we get

$$E[T_d^{\,2}] \approx \frac{D^2}{\pi R^2},\quad \mathrm{var}(T_d) \approx \left(\frac{4-\pi}{4\pi}\right)\cdot\frac{D^2}{R^2} \approx 0.068\frac{D^2}{R^2}.$$

Figures 2 and 3 show the agreement of the previous equations and the results of simulations ($10^5$ Monte Carlo runs) with $D=1$ m, $R=1$ m/s, $N$ varying from 10 to 10000 and one random ignition point.

### 3.2.2 Elliptical propagation at constant rate

In this section, we consider the case of a surface fire that propagates with an elliptical shape and at a constant rate of spread. In this case, the function $F$ is given by (see Finney (1998))

$$F(t) = \pi\cdot R^2 \frac{\left(1+\frac{1}{HB}\right)^2}{4LB}t^2,$$

where $R$ is the rate of spread, $HB$ is the Head-to-Back ratio and $LB$ is the Length-to-Breadth ratio.

In a similar fashion as before, we can compute

$$E[T_d] \approx \frac{2\sqrt{LB}}{1+\frac{1}{HB}} \frac{D}{2R}, \quad E[T_d^2] \approx \frac{4LB}{\left(1+\frac{1}{HB}\right)^2} \frac{D^2}{\pi R^2},$$

$$\text{var}(T_d) \approx \left(\frac{4-\pi}{4\pi}\right) \cdot \frac{4LB}{\left(1+\frac{1}{HB}\right)^2} \cdot \frac{D^2}{R^2} \approx 0.068 \frac{4LB}{\left(1+\frac{1}{HB}\right)^2} \cdot \frac{D^2}{R^2}.$$

## 4  Conclusions

The lack of extensive research in the application of inexpensive wireless sensor nodes for the early detection of wildfires motivated us to investigate the cost of such a network. As a first step, in this paper we present several results which relate the time to detection and the burned area to the number of sensor nodes in the region which is protected. Our main result is Theorem 1 which states that the probability distribution of the burned area is approximately exponential, given that some hypotheses hold: the positions of the sensor nodes are independent random variables uniformly distributed and the number of sensor nodes is large. It is important to remark that this conclusion depends neither on the propagation model of the fire nor on the number and distribution of ignition points.

Our next step in the investigation of the cost of a network of sensor nodes for the detection of wildfires is to actually build prototypes of such nodes and to test their behavior under controlled fires.

## Acknowledgements

This work was partially supported by anonymous contributors through the project "*Prevention and early detection of forest fires by means of sensor networks*" which is being developed at the Instituto Tecnológico de Buenos Aires (ITBA).

Figure 1

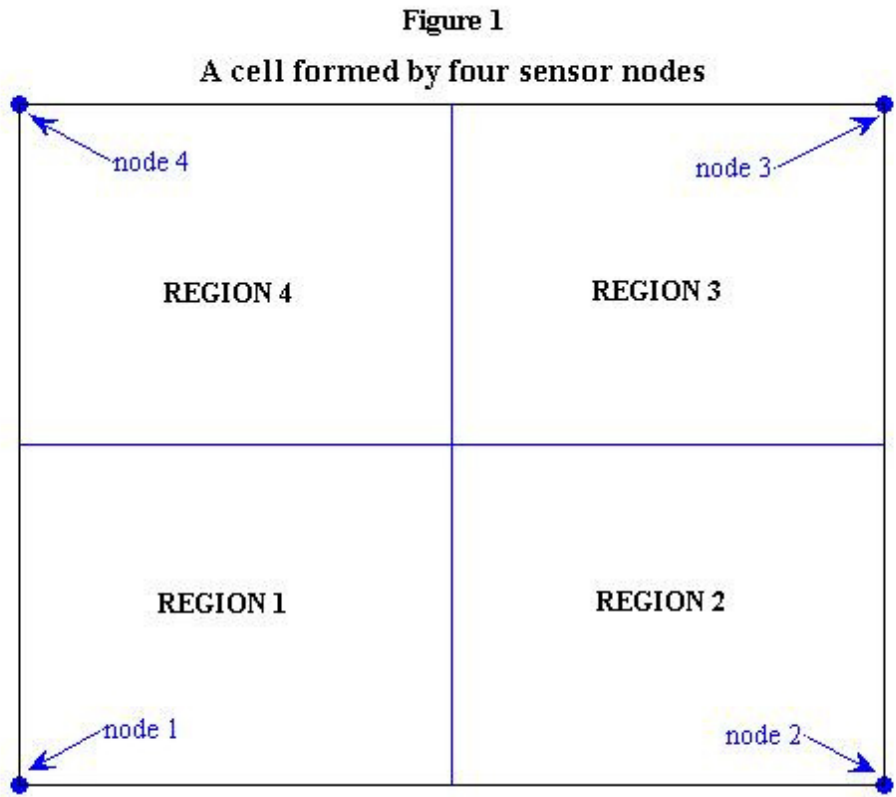

Figure 2

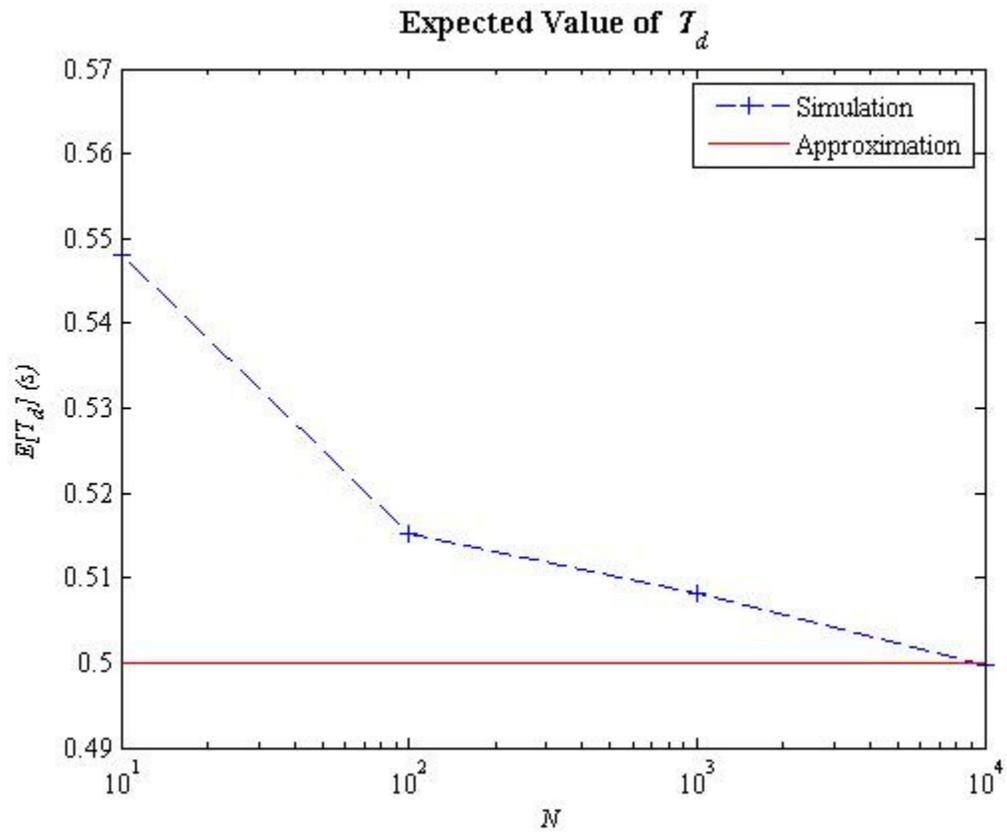

Figure 3

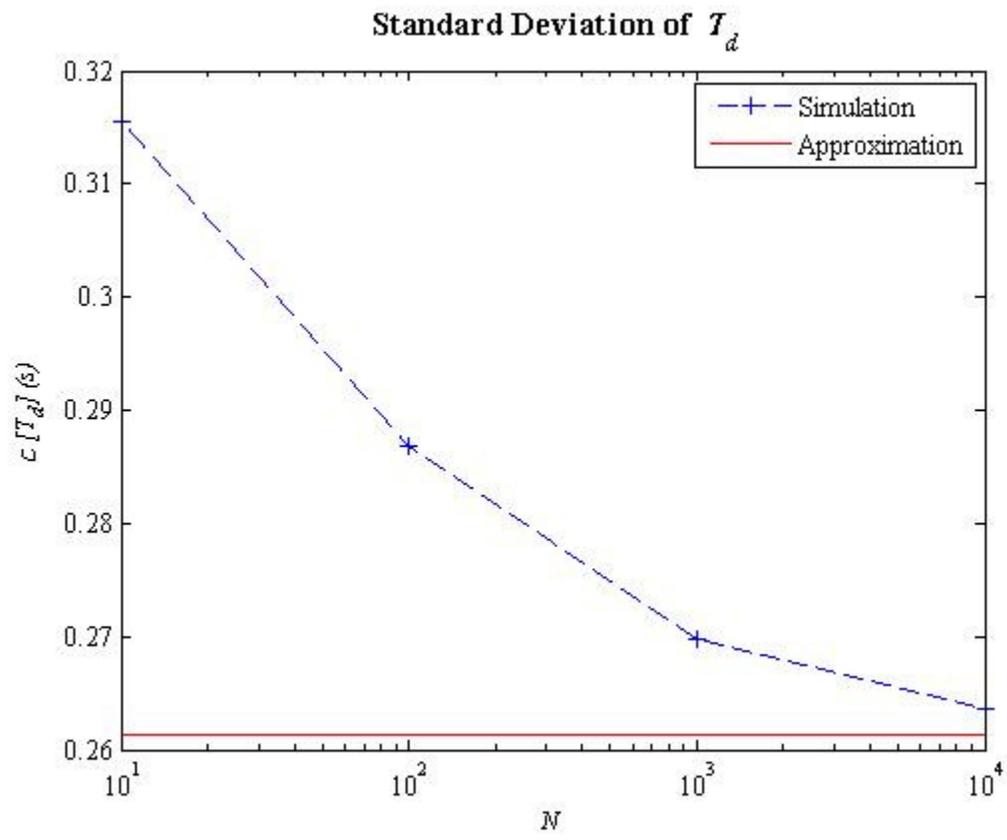